\documentclass[twocolumn,showpacs,preprintnumbers,amsmath,amssymb,prl,superscriptaddress]{revtex4}

\usepackage{graphicx}
\usepackage{dcolumn}
\usepackage{bm}

\begin{document}


\title{Series of Bulk Magnetic Phase Transitions in Na$_x$CoO$_2$: a $\mu$SR study}

\author{P.~Mendels}%
\author{D.~Bono}
\author{J.~Bobroff}
\affiliation{%
Laboratoire de Physique des Solides, UMR 8502, Universit\'e Paris-Sud, 91405 Orsay, France
}%

\author{G.~Collin}
\affiliation{
Laboratoire L\'eon Brillouin, CE Saclay, CEA-CNRS, 91191 Gif-sur-Yvette, France
}%

\author{D.~Colson}
\affiliation{
SPEC, CE Saclay, CEA-CNRS, 91191 Gif-sur-Yvette, France
}%

\author{N.~Blanchard}%
\author{H.~Alloul}
\author{I.~Mukhamedshin}
\author{F.~Bert}
\affiliation{%
Laboratoire de Physique des Solides, UMR 8502, Universit\'e Paris-Sud, 91405 Orsay, France
}%

\author{A.~Amato}%
\affiliation{%
Laboratory for Muon Spin Spectroscopy, Paul Scherrer Institute, CH-5232 Villigen PSI, Switzerland
}%

\author{A.D.~Hillier}%
\affiliation{%
ISIS Facility, Rutherford Appleton Laboratory, Chilton, Didcot, Oxon, OX11, OQX, United Kingdom
}%
\date{\today}

\begin{abstract}
Using muon spin rotation, well-defined bulk $\sim$ 100\% magnetic phases in Na$_x$CoO$_2$ are revealed. A novel magnetic phase is detected for $x=0.85$ with the highest transition temperature ever observed for $x\geq 0.75$. This stresses the diversity of $x\geq 0.75$ magnetic phases and the link between magnetic and structural degrees of freedom. For the charge-ordered $x=0.50$ compound, a cascade of transitions is observed below 85~K. From a detailed analysis of our data, we conclude that the ordered moment varies continuously with temperature and suggest that the two secondary transitions at 48~K and 29~K correspond to a moderate reorientation of antiferromagnetically coupled moments.  
\end{abstract}

\maketitle

Beyond their long-known ionic mobility which opened the route to industrial applications, cobaltates recently received  considerable attention after the discovery of high thermoelectric power in metallic Na$_{0.7}$CoO$_2$~\cite{Terasaki} and superconductivity, maybe unconventional, in the hydrated Na$_{0.35}$CoO$_2$ compound~\cite{Takada}. In addition, a very rich phase diagram~\cite{Cava}, still to be explored in detail, seems to involve many and possibly competing parameters such as doping, charge order, magnetism, frustration and strong electronic correlations. This spans over most of topical problems in condensed matter, more specifically in the field of correlated systems. Whether one parameter has a leading role over others is a central issue for understanding the fundamentals of physics in Na-cobaltates.

Cobaltates are layered compounds, like high-$T_c$ cuprates. Magnetic and conducting properties set in CoO$_2$ layers, issued from edge sharing CoO$_6$ octahedra stacked along the $c$-axis and separated by Na$^+$ partially filled layers. In a naive model, the increase of Na$^+$ content leads to a conversion of the formal charge of Co from 4$^+$ (low-spin, $S=1/2$) to 3$^+$ ($S=0$) -hence a depleted antiferromagnetic frustrated quantum triangular lattice- and/or to electron doping occuring in a $S=1/2$ background, which \emph{per se} are interesting problems. 

Experimentally, both metallic and magnetic characters can be found for various $x$ but, at variance with high-$T_c$'s, Na$^+$ and Co$^{3+/4+}$ charge-orderings might set in for well defined compositions as shown by recent electronic diffraction~\cite{Zandbergen}, cristallographic~\cite{Huang} and NMR studies~\cite{Mukhamedshin}. 

Magnetic states have been reported for long for $x\geq 0.75$. A commensurate spin density wave (C-SDW) was detected below 22~K~ in minority volume fraction in a nominal $x=0.75$ sample~\cite{Sugiyama}. Given the reported $c$-axis parameter is rather typical of the non-magnetic $x=0.70$ phase~\cite{Mukhamedshin}, the invoked magnetic inhomogeneity might rather relate to the multiphasing of the investigated compound.  Furthermore, on the basis of a second nominally $x=0.90$ sample where an incommensurate (IC)SDW 25-50\% fraction order sets in below 20~K, a dome-shaped continuous phase diagram was proposed, in a Hubbard approach which links magnetism to doping~\cite{Sugiyamabis}. \emph{Bulk} C-SDW was reported to occur below 20~K~\cite{Stuttgart}, only in \emph{one} $x=0.82(2)$ single crystal, with a $\mu$SR signature different from the previous samples. Sorting out the actual impact of doping on magnetism of the CoO$_2$ planes clearly calls for a more refined study.  

The importance of the structure might be well illustrated in the peculiar $x=0.50$ case. A well defined commensurate superstructure has been observed~\cite{Zandbergen} where Na$^+$/vacancies and Co$^{3+}$/Co$^{4+}$ arrange in order to minimize both Na-Na and Na-Co Coulomb repulsion. A magnetic state was recently reported below 50~K~\cite{Uemura} and a metal-insulator transition sets in around 30~K~\cite{Cava}. In addition a susceptibilty anomaly is observed around 90~K~\cite{Cava}, which origin is still not known.

In this Letter, we present a study of $\sim100\%$ magnetic phases, detected through $\mu$SR. The structures of all investigated samples were checked by room-$T$ X-ray Rietveld refinements. For $x\geq0.75$, we isolate two \emph{pure} magnetic phases, including a novel one for $x=0.85$. We clearly demonstrate the absence of phase separation, stress the underlying role of the structure , hence the complexity of the magnetic, likely non-continuous phase diagram. For $x=0.50$, we clearly establish, for the first time, that 3 magnetic transitions occur below 85~K.  

All samples were prepared by solid state reaction of Co$_3$O$_4$ and Na$_2$CO$_3$. They were further annealed in air for twice 12 hrs and quenched with intermediate grinding, then annealed in flowing oxygen for 24 hrs at 600$^\circ$C.   
The $x=0.50$ composition was reached by immerging and stirring for 4 days the starting $x=0.70$ material in a sodium hypochlorite solution. Hexagonal parameters $a=2.81511(3)$~\AA\ and $c=11.1314(2)$~\AA\ were measured with a $Pnmm$ orthorhombic superstructure ($a\sqrt{3},2a,c$). Na sites, at the vertical either of a Co site (Na1) or of the center of a Co triangle (Na2) have an equal occupancy of 0.25, in agreement with the nominal composition.
The $x=0.75$ sample displays an hexagonal $P6_3$/$mmc$ lattice with $a=2.84175(4)$~\AA\ , $c=10.8087(2)$~\AA\ , typical of the H2 phase of ref.~\cite{Huang} and exhibits only a few very weak (incommensurate) additional diffraction peaks for $25<2\theta<40^\circ$. Structural refinements lead to site occupancies 0.22(1) for Na1 and 0.52(1) for Na2, close to the nominal content 0.75. This material decomposes spontaneously in air leading to the $x=0.70$ typical diffraction pattern ($c \sim 10.892(4)$~\AA\ ). Finally, $x=0.85$ material exhibits a monoclinic distortion with a 10.766(2)~\AA\ $c$-lattice constant (projected on the hexagonal cell).

The $\mu$SR experiments were performed at the ISIS (EMU) and PSI (GPS) facilities~\footnote{Lose powders or pressed disks were used. Due to a more or less pronounced platelet shape of the grains, some c-axis preferential orientation could be observed $\sim$ parallel to the $\mu^+$ polarization}. In the case of magnetic order, spontaneous oscillations of the muon polarization are detected without any applied field (ZF setup) through the asymmetry of the positron emission due to the $\mu^+$ desintegration. The frequency (ies) correspond to the internal field(s)($H_{int}$) at the muon(s) site(s) and, in the case of simple magnetic phases, track the variation of the order parameter (\emph{i.e.}~the local static moment). High statistics counts allowed to reveal clearly \emph{all} the oscillating frequencies for each detected magnetic phase.
   
\begin{figure}[tb] \center
\includegraphics[width=0.85\linewidth]{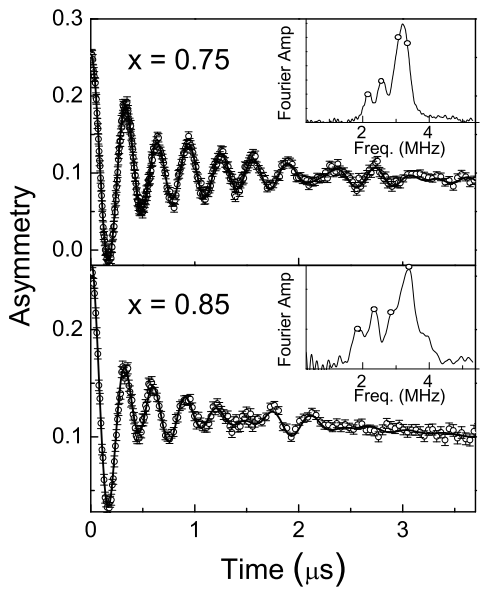}
      \caption{ \label{fig_asyTF0p75}
ZF asymmetries (lines are for fits) and FT at 5~K for $x=0.75$ and 0.85 (PSI). Circles on FT are for fitted frequencies The asymmetry of the high frequency line is accounted for by 2 frequencies (\emph{i.e.} 2 $\mu^+$ sites) for $x=0.75$. 
       }
      \end{figure}
      
We first give a brief account for the $x\geq 0.75$ samples which display a textbook $\mu$SR signature of a phase transition to an ordered magnetic state. In Fig.~\ref{fig_asyTF0p75}, for $x=0.75$ and $0.85$ samples, we report both ZF asymmetry and Fourier transforms (FT) which give the distribution of $H_{int}$ over all $\mu^+$ sites.  The FTs are quite well peaked clearly indicating a simple magnetic order in view of the number of $\mu^+$ sites expected (see $x=0.5$ discussion).
Accordingly, $\mu$SR signals and their $T$-evolution are best described by damped cosine functions (see ref.~\cite{Sugiyama,Stuttgart} for details), the weight of which can be fixed at low $T$ values.

 \begin{figure}[tb] \center
\includegraphics[width=1\linewidth]{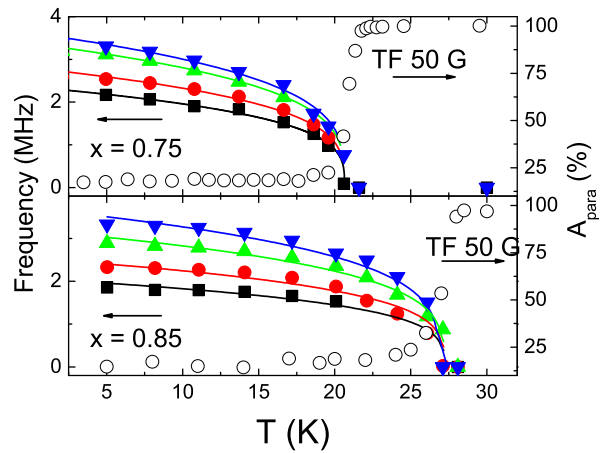}
      \caption{ \label{fig_freq0p75fits}
Left, closed symbols: $T$- variation of the 4 ZF frequencies used in the fits. $\sim$ $(T-T_o)^{0.28}$ lines emphasize the scaling of all their variations. The widths are smaller or of the order of the point sizes. Right, open circles: wTF asymmetry .}
      \end{figure}
      
In Fig.~\ref{fig_freq0p75fits}, we plot the weak transverse field (wTF) asymmetry which monitors the non-magnetic fraction, found to be less than 10\% in both samples. The sharpness of the decrease of the wTF signal allows us to extract transition temperatures, $T_o=$ 20.8(5) and 27(1)~K, for $x=0.75$ and 0.85. For each sample, the frequencies (Fig.~\ref{fig_freq0p75fits}) scale with each other on the whole $T$-range and decrease smoothly when $T$ increases to vanish at $T_o$, simply reflecting the variation of the order parameter of a unique magnetic phase probed at 4 different $\mu^+$ sites. 

For $x= 0.75$, the frequencies and $T_o$ values are similar to that of minority $\sim$20\% phases reported in~\cite{Sugiyama}. On the contrary, to our knowledge, for $x = 0.85$, the frequencies are different from existing ones and the transition temperature is much higher than any reported to date for $x \ge 0.75$. This points at a novel magnetic phase, commensurate, as suggested above, which is worth noticing for such non-peculiar value of $x$ as 0.85. 

\begin{figure}[tb] \center
\includegraphics[width=1\linewidth]{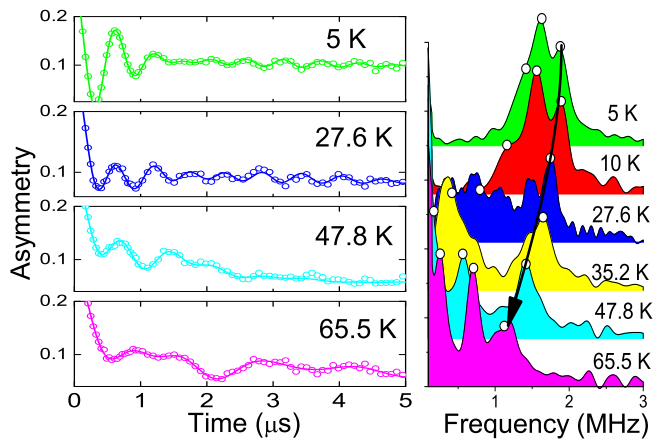}
      \caption{ \label{fig_asyTF0p5}
Left panel: ZF asymmetries for $x=0.5$ at various typical~$T$ (PSI). Lines are fits, see  text. The evident change of the oscillations shape is reflected in the Fourier transforms (right panel). There, circles mark fitted frequencies.
		}
      \end{figure}
      
\begin{figure}[tb] \center
\includegraphics[width=1\linewidth]{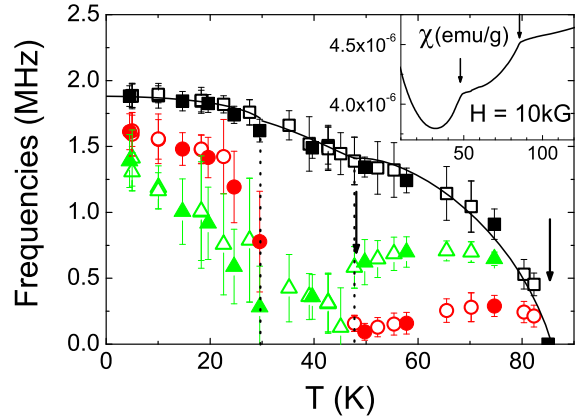}
      \caption{ \label{Fig_freqfit}
$T$-plot of the $x=0.5$ ZF frequencies (the line is a guide to the eye). Bars stand for the FWHM of the various spectral lines. The errors in the fits are smaller than the data points. Closed (open) symbols are for PSI (ISIS) data taken on different batches. Vertical arrows indicate the kinks of the magnetic susceptibility (see inset). Vertical dotted lines separate the various regimes. 
       }
      \end{figure}

We now focus on the $x=0.50$ composition. In Fig.~\ref{fig_asyTF0p5}, we display for various $T$ both asymmetries versus time and their FT. Magnetic order is evident through spontaneous oscillations of the ZF $\mu$SR signals below 85(1)~K, which corresponds to the high $T$ kink in SQUID data (inset, Fig.~\ref{Fig_freqfit}). Unlike previous cases where the shape of the asymmetry remains similar for all $T$ in the frozen regime, we find, for the first time, that \emph{three} distinct magnetic regions exist for $x=0.50$~\footnote{For $T>48$~K, no magnetism was found in [\onlinecite{Uemura}]}. Depending on the $T$ range, 2 or 3 frequencies  were used to fit the $\mu$SR data, leading to a more complex $T$-dependence of the frequencies (Fig.~\ref{Fig_freqfit}) than in Fig.~\ref{fig_freq0p75fits}. Two extra-transitions are singled out at lower $T$, namely at 29(1)and 48.0(5)~K.  

At base temperature (5~K), three frequencies can be distinguished on the FT in the range 1.3-1.9 MHz. Surprisingly, upon increasing the temperature, only the high frequency line is kept, whereas the remaining spectrum  progressively spreads downwards \emph{continuously} to even reach zero frequency around 29~K. Correlatively, the widths of the two lowest frequency signals noticeably increase from $T=5$~K to finally overlap for $T\lesssim 28$~K. Between 30 and 48 K, the FT spectrum is mainly peaked at two frequencies, one  fairly low ($<0.5$~MHz) and one in the continuity of the high frequency cutoff. The FT spectrum changes abruptly around 48~K and three frequency peaks are again evident between 50 and 85~K. 

If the highest-frequency still roughly mimics the variation expected for an order parameter, the variation of the other frequencies do not scale with this one and have an unconventional non-monotonic variation in the $T>48$~K phase. The former certainly corresponds to a $\mu^{+}$ dominantly coupled to a single magnetic site while the latter are issued from several moments and are likely sensitive to the relative orientation of moments. In this respect, the continuity of the highest frequency indicates that there is no sudden change of the moment at 48 and 29~K. On the contrary, the marked change at 48~K for the others reveals a lock-in of the magnetic structure below 48~K -and down to 29~K- which is confirmed by the kink observed in SQUID data at 48~K. wTF measurements (not displayed) indicate that the magnetic fraction in our samples is $\sim$ 90\% for all $T<85$~K, and therefore underlines the bulk character at all $T$. Finally, the changes observed for \emph{all} $\mu^+$ frequencies at these specific $T$ either through inflexions or discontinuities clearly establish that we deal, in each domain, with a single cristallographic phase and no phase separation occurs.

In order to further discuss our data, we now address the issue of the $\mu^+$ location. Possible sites, minimizing the $\mu^+$ electrostatic energy are either a Na$^+$ vacancy or a site located 1~\AA\	away from an O$^{2-}$, forming a O-$\mu^-$ bond. The first one can be discarded since (i) nearby Na$^+$ would repel the $\mu^+$ from this position (ii) the Gaussian damping in the paramagnetic state, originating from Na and Co nuclear dipoles,  is much too small to match with this site location \emph{only} ($\Delta_{calc}=0.28(1)$ instead of 0.17 $\mu$s$^{-1}$) and (iii) in the case of \emph{partial} occupancy, a much lower frequency than observed at 5~K should be found at a Na vacant site in comparison with that at O$^{2-}$ site (ratio $\sim$ 1:5), as already advocated in~\cite{Stuttgart,Sugiyama} for $x\geq 0.75$. 

\begin{figure}[b] \center
\includegraphics[width=0.5\linewidth]{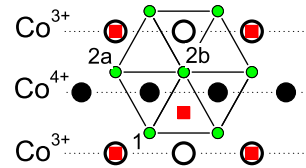}
      \caption{ \label{Fig_structure}
Charge order suggested in ref.\onlinecite{Zandbergen}. Big circles are for Co$^{4+/3+}$ and squares are for Na above the Co plane. Oxygens (small circles) form a triangular network shifted from the Co one above and below (not represented) the Co planes.
       }
      \end{figure}

For $x=0.50$, Co are known to form alternating chains in a given plane \cite{Zandbergen}. Every 2 chains, Na alternate above and below each Co, hence the Co valence in such a chain is expected to be uniform and weaker than for the next Co, also uniform, chain. In the following, we make the simple assumption that cobalt is either in the Co$^{3+}$ or Co$^{4+}$ state~\footnote{Whatever the refinements about the charge and magnetic state of Co, we expect our discussion to hold provided only two dominant Co species exist for $x=0.5$, one strongly magnetic and the other weakly magnetic.} and consider an ideal undistorted orthorhombic structure. We arbitrarily consider $\mu^+$ bound to oxygens located above the Co plane. Inequivalent charge/magnetic configuration surrounding oxygens can be sorted by considering Co triangles next to the $\mu^+$ site altogether with the occupancy of \emph{nn} Na$^+$ sites (fig.~\ref{Fig_structure}). Three inequivalent O$^{2-}$ are found, consistent with the number of frequencies observed at 5~K and above 49~K. Two have 2 \emph{nn} magnetic Co$^{4+}$ and have either a \emph{nn} Na1 or a \emph{nn} Na2, respectively labeled (2a) and (2b)~\footnote{Due to the symmetry of the charge distribution, the $\mu^+$ locates in the mid-plane of the two \emph{nn} Co$^{4+}$ segment.}. The third site (1) has a single \emph{nn} Co$^{4+}$ and 2 \emph{nn} Co$^{3+}$ with one Na1 and one Na2 filled nearby sites. 

As already argued above, $\mu^+$ with 1 magnetic \emph{nn} Co$^{4+}$ (site 1) can be safely assigned to the highest frequency signal. A $\mu^+$ coupled to two \emph{nn} Co$^{4+}$ will sense pronounced changes of internal fields and may also experience quite weak fields. Our data constrains severely the orientation of the moments, since, first, $H_{int}$ is observed to be weaker for sites (2) than for sites (1) for \emph{all} $T$ and second, the ratio varies sizeably with $T$ to reach $\sim$ 1/5 value between 29 and 48~K. To illustrate this point, one can notice that $H_{int}$(2) cancels only for two antiferromagnetic Co$^{4+}$ moments, lying perpendicular to the Co$^{4+}$-Co$^{4+}$-$\mu^+$ plane (hence perpendicular to the chain).
       
We first performed calculations of the dipolar field induced on the 2 types of sites by a chain of Co$^{4+}$, assuming an AF coupling. The effect of moments beyond 2$^{nd}$ \emph{nn} was found negligible. We further checked that the drawn conclusions still hold without sensible changes for a 3D magnetic stacking either ferro or antiferromagnetic of such AF chains. Our major findings are (i) the moments lie close to AF order (ii) their direction is within 20~$^\circ$~ perpendicular to the plane O(2)-\emph{nn} Co$^{4+}$ chain; (iii) sites (2) frequencies are fairly sensitive to a small misalignment of neighbor moments or to a change of the direction of the AF ordered moments, \emph{e.g.}  10-20$^\circ$ is enough to explain the variations observed in Fig.~\ref{Fig_freqfit}. Note that the moment direction should alternate along $c$-axis since oxygen triangles are inverted for two next Co planes. 

For this structure, since site (1) is coupled dominantly to a single Co, the highest frequency at 5~K can be used to extract a value for the Co$^{4+}$ moment of the order of 0.3(1)~$\mu_B$.
Such a weak value could be explained by a strong zero point quantum reduction due to the possible 1D character of the Co$^{4+}$ chain. Alternatively, a commensurate spin density wave description in the framework of band magnetism -the system is metallic above 29~K- could be a natural explanation of this value.  

The values of the frequencies are found to be only little affected by a change in the $\mu^+$ position as compared to the effect of moments misalignement. Therefore, the changes around 29~K cannot be merely explained by a slight structural change or a charge re-ordering which would affect the $\mu^+$ site. We infer that a magnetic component is also associated with this transition, smoother than the 48~K one. The increase of the resistivity observed in~\cite{Cava}, rather modest in comparison with usual metal-insulator transitions, clearly shows that electronic and magnetic properties have a common origin, likely structural.

The origin of these multiple transitions is not obvious at the present stage. We can clearly rule out a scenario where an IC-SDW would switch to a C-SDW since the frequencies are already quite well peaked at the upper transition. Whether the problem can be tackled through a purely ionic model where anisotropy and charge localization could induce the secondary transitions or whether a C-SDW order might be influenced by these parameters open new avenues to the debate on this peculiar $x=0.5$ composition.         

In summary, we clearly show the existence of bulk phase transitions in high quality powdered samples with well identified structures. This truly underlines the importance of magnetism in cobaltates, in a large range of Na doping. For $x=0.5$, we give a convincing evidence that three magnetic transitions occur with an AF order building up at 85~K and moments rearrangements at lower $T$ (48 and 29~K). For $0.75\leq x\leq 0.85$, one can now isolate three close magnetic phases, with a C-SDW ground state, including the $x=0.82$ of~\cite{Stuttgart}~\footnote{During the refereeing rounds, we also isolated this phase}. The increase of $T_o$ from 21~K for the hexagonal $x=0.75$ to 27~K for the monoclinic $x=0.85$ phase is quite surprising since the number of magnetic Co is expected to decrease. Whether Co-Co couplings differ because of structural changes, or some charge ordering locks-in a different magnetic order definitely calls for further investigations of stable structures in the range $x=0.75-1$ as well as a clear identification of magnetic Co sites with respect to the Na order. Rather than a continuous phase diagram induced by charge doping only, the scattering of the $T_o$s rather point at a tight link between structures and magnetic order. Disentangling how charge order/disorder and doping impact on the physical properties versus Na content is now a crucial issue for cobaltates which apparently combine the physics of manganates and high-$T_c$ cuprates.

\end{document}